\documentclass[aps,prl,twocolumn,superscriptaddress,showpacs]{revtex4}
\usepackage{keyval}%
\usepackage{graphicx}
\usepackage{dcolumn}
\usepackage{bm}
\setkeys{Gin}{width=0.8\columnwidth,clip=true}

\begin{document}

\title{Spectral functions of the Falicov-Kimball model with electronic ferroelectricity}

\author{Wei-Guo Yin}
\email{wyin@mail.unomaha.edu}
\author{W. N. Mei}%
\author{Chun-Gang Duan}%
\affiliation{%
Department of Physics, University of Nebraska, Omaha, NE 68182}
\author{Hai-Qing Lin}
\affiliation{Department of Physics, The Chinese University of Hong
Kong, Hong Kong, People's Republic of China}
\author{J. R. Hardy}
\affiliation{Department of Physics and Center for Electro-Optics,
University of Nebraska, Lincoln, NE 68588}%


\date{\today}

\begin{abstract}
We calculate the angular resolved photoemission spectrum of the
Falicov-Kimball model with electronic ferroelectricity where $d$-
and $f$-electrons have different hoppings. In mix-valence regimes,
the presence of strong scattering processes between $d$-$f$
excitons and a hole, created by emission of an electron, leads to
the formation of pseudospin polarons and novel electronic
structures with bandwidth scaling with that of $d$-$f$ excitons.
Especially, in the two-dimensional case, we find that flat regions
exist near the bottom of the quasiparticle band in a wide range of
the $d$- and $f$-level energy difference.

\end{abstract}



\maketitle



Ferroelectric materials have long been important to physical
research and technological application.
Besides commonly known displacive and order-disorder mechanisms
\cite{efe:kittel}, recently the idea of electronic
ferroelectricity (EFE) has been attracting considerable attention.
EFE was introduced by Portengen {\em et al.} \cite
{efe:portengen,efe:portengen-prb} in their mean field theory of an
extended Falicov-Kimball model (FKM) \cite{efe:FK}. The FKM,
having a long successful history in dealing with correlated
electron systems, was then used to model a system consisting of
two bands of different parity, say itinerant $d$ electrons and
localized $f$ orbitals, in which mix-valence states may occur
depending on the $d$-$f$ coupling strength. Portengen {\em et al.%
} pointed out that inclusion of a $d$-%
$f$ hybridization in the FKM could give rise to a spontaneous
electric polarization due to a Bose-Einstein condensation (BEC) of
$d$-$f$ excitons when the excitation energy goes to zero at the critical value of the $f$%
-level energy \cite{efe:portengen,efe:portengen-prb}. It is
expected that such a purely electronic mechanism would provide
fascinating physical features which are desirable for many
applications; for instance, the static dielectric constant in an
electronic
ferroelectric could exceed $10^4$ 
\cite{efe:portengen-prb}.

However, EFE in the FKM is still controversial after tested by
different theoretical treatments of the underlying strongly
correlated electron system
\cite{efe:czycholl,efe:farkasovsky-02,efe:zlatic}. Most recently,
Batista proposed a new extension of the FKM in which an $f$-$f$
hopping was included \cite {efe:batista}. Mapping the strong
coupling limit of this model into an $xxz$ pseudospin $1/2$ model
with a magnetic field along the $z$-axis and supported by quantum
Monte Carlo calculations performed earlier by Schmid \textsl{el
al.} \cite{efe:schmid}, he showed that a BEC of $d$-$f$ excitons
does exist in the phase diagram \cite {efe:batista}.

The purpose of this Letter is to present a theoretical study of
the angular resolved photoemission spectrum of this extended FKM.
In the pseudospin picture \cite{efe:batista}, we study the
spectral functions of one hole, created by emission of one
electron, in an extended $t$-$J$ model with different hoppings for
each pseudospin flavor. The problem of single hole motion in a
local (pseudo)spin background has become an essential issue in
understanding anomalous physical properties of high temperature
superconductors \cite{HTC:dagotto2,HTC:SVR,yin:prl98} and of
colossal magnetoresistance manganites \cite{yin:prl01,CMR:bala}.
It has been demonstrated that quantum antiferromagnetic
(pseudo)spin fluctuations have strong impact on the low-energy
scale physics of these materials. In this Letter, we show that
strong scattering processes between the hole and the $d$-$f$
excitons take place even in the ferroelectric regime that has a
\emph{ferromagnetic} pseudospin configuration, thus give rise to
the formation of pseudospin polarons and novel electronic
structures.

The extended Falicov-Kimball model for spinless fermions on a
hypercubic lattice is \cite{efe:batista}:
\begin{equation}
H=\epsilon _d\sum_\mathbf{i}n_\mathbf{i}^d+\epsilon
_f\sum_\mathbf{i}n_\mathbf{i}^f+t_d\sum_{\langle
\mathbf{i},\mathbf{j}\rangle }d_\mathbf{i}^{\dagger
}d_\mathbf{j}^{}+t_f\sum_{\langle \mathbf{i},\mathbf{j}\rangle
}f_\mathbf{i}^{\dagger
}f_\mathbf{j}^{}\,+U^{fd}\sum_\mathbf{i}n_\mathbf{i}^dn_\mathbf{i}^f,
\label{H0}
\end{equation}
where $n_\mathbf{i}^d=d_\mathbf{i}^{\dagger }d_\mathbf{i}^{}$ and
$n_\mathbf{i}^f=f_\mathbf{i}^{\dagger }f_\mathbf{i}^{}$ are the
occupation numbers of $d$- and $f$-orbitals, respectively.
Representing the
two orbital flavors by a spin $1/2$ variable, $c_{\mathbf{i}\uparrow }=d_\mathbf{i}$ and $%
c_{\mathbf{i}\downarrow }=f_\mathbf{i}$, and using the notion of the pseudospin operator %
$\mathbf{\vec{\tau}}_\mathbf{i}=\sum_{\mu \nu }c_{\mathbf{i}\mu
}^{\dagger }\mathbf{ \vec{\sigma} }^{}_{\mu \nu }c_{\mathbf{i}\nu
}^{}$ with $\{\mathbf{ \vec{\sigma} }^{}_{\mu \nu }\}$ being the
Pauli matrices, one can derive an effective $t$-$J$ model,
$H_{{\rm eff}}=H_t+H_J$, in the strong coupling limit of
(\ref{H0}) near half-filling \cite{efe:batista}:
\begin{eqnarray}
H_J &=&\sum_{\langle \mathbf{i},\mathbf{j}\rangle }[J_z\tau
_\mathbf{i}^z\tau _\mathbf{j}^z+J_{\perp }(\tau _\mathbf{i}^x\tau
_\mathbf{j}^x+\tau
_\mathbf{i}^y\tau _\mathbf{j}^y)]+B_z\sum_\mathbf{i}\tau _\mathbf{i}^z,\nonumber \\
H_t &=&\sum_{\langle \mathbf{i},\mathbf{j}\rangle ,\sigma
}t_\sigma \,(\widetilde{c}_{\mathbf{i}\sigma
}^{\dagger }\widetilde{c}_{\mathbf{j}\sigma }^{}\,+\widetilde{c}_{\mathbf{j}\sigma }^{\dagger }%
\widetilde{c}_{\mathbf{i}\sigma }^{}),
\end{eqnarray}
where $\widetilde{c}_{\mathbf{i}\sigma }^{}=c_{\mathbf{i}\sigma
}^{}(1-n_{\mathbf{i}\overline{\sigma }}) $ is the constrained
fermion operator, $t_{\uparrow }=t_d$, $t_{\downarrow }=t_f$ ,
$J_z=2(t_{\uparrow}^2+t_{\downarrow}^2)/U^{fd}$, $J_{\perp
}=4t_{\uparrow }t_{\downarrow
}/U^{fd}$, and $%
B_z=\epsilon _d-\epsilon _f$ is the $d$- and $f$-level energy
difference acing as a magnetic field along the $z$ direction.

\begin{figure}
\includegraphics*{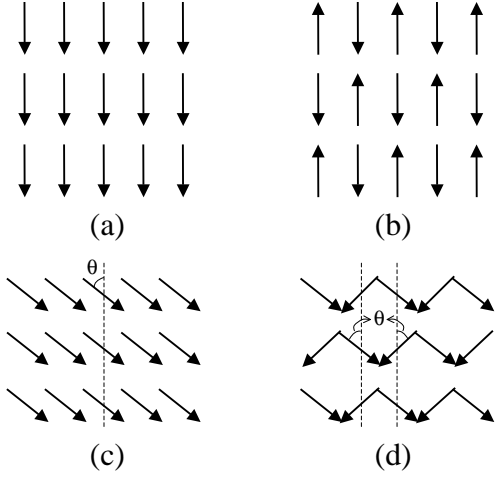}%
\caption{\label{fig:phases}%
Schematic pictures of the pseudospin configuration in (a) full
$f$-band, (b) staggered orbital ordering, (c) ferroelectric
canting state, and (d) antiferroelectric canting state.}
\end{figure}

\begin{figure}
\includegraphics*{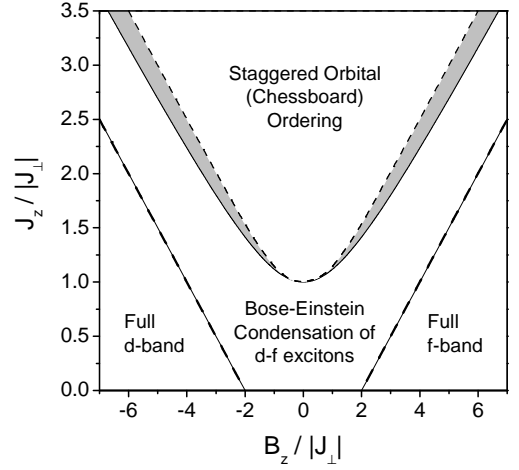}%
\caption{\label{fig:phasediagram}%
Two dimensional phase diagram of $H_J$ obtained from linear spin
wave theory (solid lines) and from quantum Monte Carlo
calculations [\onlinecite{efe:batista,efe:schmid}] (dashed lines).
Shadow area denotes possible phase separation.
}
\end{figure}

At half filling only $H_J$ contributes to $H_\mathrm{eff}$. The
ground state of $H_J$ is one of the following phases:
full $d$-band, full $f$-band, staggered orbital ordering, and the
BEC of $d$-$f$ excitons that is ferroelectric for $J_{\perp }<0$
or antiferroelectric for $J_{\perp }>0$ \cite {efe:batista}. We
assume that the BEC of $d$-$f$ excitons is single-phased -- in the
present case, the corresponding pseudospin configuration is a
canting state with respect to the $z$ axis
[Fig.~\ref{fig:phases}(c) or \ref{fig:phases}(d)], where the
canting angle
$\theta $ is determined by %
\begin{equation}
\cos \theta =\frac{|B_z|}{zS(J_z+|J_{\perp }|)},
\end{equation}
where $S=1/2$ is the value of the pseudospins and $z$ is the
coordination number. Indeed, the spin excitation (i.e., $d$-$f$
exciton) in the canting state is a gapless Goldstone mode which
corresponds to a uniform precession of the pseudospins around the
$z$ axis. Now we diagonalize $H_J$ in linear spin wave theory and
obtain the following phase diagram as shown in
Fig.~\ref{fig:phasediagram}: The borderline between staggered
orbital ordering and the BEC of $d$-$f$ excitons is
$|B_z|=zS\sqrt{J_z^2-J_{\perp }^2}$ and the corresponding phase
transition is a first-order transition; the borderline between the
BEC of $d$-$f$ excitons and full $d$- or $f$- band is
$|B_z|=zS(J_z+|J_{\perp }|)$ and the corresponding transition is a
second-order transition.
In a whole, our results agree with the quantum Monte Carlo studies
on finite size systems up to $96\times 96$ \cite{efe:schmid}.
Notice that in Fig.~\ref{fig:phasediagram} the borderline between
staggered orbital ordering and the BEC of $d$-$f$ excitons
obtained from linear spin wave theory (solid line) does not
perfectly coincide with that from the Monte Carlo calculations
(dashed line). This might result from our single phase
approximation, indicating the existence of a phase separation
regime between these two lines (shadow area). Nevertheless, the
difference is small if the values of $J_z$ and $|J_\perp|$ are
close to each other, and this is favorable for ferroelectricity to
take place \cite{efe:batista}. We shall analyze the hole dynamics
in each phase in the following.

{\em Full }$f${\em -band} \cite{note:phases}. The pseudospin
configuration is a ferromagnetic state
with all pseudospins down [Fig~\ref{fig:phases}(a)]. The pseudospin wave spectrum is $\omega _{{\bf q}%
}^{full}=B_z-zS(J_z+|J_{\perp }|\gamma _{{\bf q}})$ where $\gamma _{{\bf q}%
}=\sum_{\mathbf{\delta }}e^{i{\bf q}\cdot \mathbf{\delta}}/z$ and
$\mathbf{\delta }$ is a unit vector connecting nearest neighbors.
Since the hole motion will not change the pseudospin
configuration, the hole can move freely with dispersion
$\varepsilon _{{\bf k}}^{full}=-zt_{\downarrow }\gamma _{{\bf
k}}$.

{\em Ferroelectric phase.} The pseudospin configuration is
considered here as a uniform canting state
[Fig~\ref{fig:phases}(c)]. The pseudospins are parallel, but they
take new equilibrium direction which are tilted by a certain angle
$\theta$ to the $z$ axis. This new axis of quantization is
considered to lie in the $x$-$z$ plane as a result of spontaneous
symmetry breaking. We perform a uniform rotation of the orbitals
about the $y$ axis by $\theta $,
\begin{eqnarray}
\widetilde{c}_{{\bf i}\uparrow } &=&\cos \frac \theta 2e_{{\bf
i}\uparrow
}-\sin \frac \theta 2e_{{\bf i}\downarrow },\;\widetilde{c}_{{\bf i}%
\downarrow }=\sin \frac \theta 2e_{{\bf i}\uparrow }+\cos \frac \theta 2e_{%
{\bf i}\downarrow },  \nonumber \\
\tau _\mathbf{i}^x &=&S_\mathbf{i}^x\cos \theta
-S_\mathbf{i}^z\sin \theta ,\;\tau
_\mathbf{i}^z=S_\mathbf{i}^z\cos \theta +S_\mathbf{i}^x\sin
\theta,
\end{eqnarray}
to obtain the ferromagnetic configuration $|\cdots S\,_{{\bf i}}^zS\,_{{\bf i%
}+1}^z\cdots \rangle =|\cdots \downarrow \downarrow \cdots \rangle
$ as the vacuum state. Then, we employ the slave-fermion formalism
to cope with the constraint of no doubly occupancy \cite{HTC:SVR}.
Defining holon (spinless fermion) operators $h_{%
{\bf i}}$ so that $e_{{\bf i}\uparrow }=h\,_{{\bf i}}^{\dagger }a_{{\bf i}%
}^{}$,$\,\,e_{{\bf i}\downarrow }=h_{{\bf i}}^{\dagger }$ where $a_{{\bf i}%
}=S\,_{{\bf i}}^{-}$ is the hard-core boson operator, we arrive at
an effective pseudospin-polaron Hamiltonian in the momentum space
\begin{eqnarray}
H_t &=&\sum_{{\bf k}}\varepsilon _{{\bf k}}^{}h_{{\bf k}}^{\dagger }h_{{\bf k%
}}^{}+\sum_{{\bf k},{\bf q}}M_{{\bf kq}}h_{{\bf k}}^{\dagger }h_{{\bf k}-%
{\bf q}}^{}\alpha _{{\bf q}}^{}+H.c.,
\end{eqnarray}
where $\alpha _{{\bf q}}$'s are pseudospin wave operators, $a_{{\bf q}}=u_{{\bf %
q}}\alpha _{{\bf q}}+v_{{\bf q}}\alpha _{-{\bf q}}^{\dagger }$,
$\,$with dispersion $\omega _{{\bf q}}^{}=zS(I_{\theta ,{\bf
q}}^2-C_\theta ^2\gamma
_{{\bf q}}^2)^{1/2}$. The transformation coefficients are $u_{{\bf q}%
}=\{[I_{\theta ,{\bf q}}(I_{\theta ,{\bf q}}^2-C_\theta ^2\gamma _{{\bf q}%
}^2)^{-1/2}+1]/2\}^{1/2}$ and $v_{{\bf q}}=-{\rm sgn}(C_\theta \gamma _{{\bf q}%
})\{[I_{\theta ,{\bf q}}(I_{\theta ,{\bf q}}^2-C_\theta ^2\gamma _{{\bf q}%
}^2)^{-1/2}-1]/2\}^{1/2}$.
Here the shorthand notations are
$B_\theta=(-|J_{\perp }|\cos ^2\theta +J_z\sin ^2\theta -|J_{\perp%
}|)/2$, $C_\theta =B_\theta +|J_{\perp }|$, and $I_{\theta ,{\bf
q}}=|J_{\perp }| + B_\theta \gamma _{{\bf q}}$. Hence $\omega
_{{\bf q=0}}=0$.
The bare hole dispersion is $%
\varepsilon _{{\bf k}}^{}=-z(t_{\downarrow }\cos ^2\frac \theta 2%
+t_{\uparrow }\sin ^2\frac \theta 2)\gamma _{{\bf k}}$, and the
hole-pseudospin-wave coupling is
\begin{equation}
M_{{\bf kq}}=\sin \theta \frac{t_{\uparrow }-t_{\downarrow }}2\frac z{\sqrt{N%
}}(\gamma _{{\bf k}-{\bf q}}u_{{\bf q}}+\gamma _{{\bf k}}v_{{\bf
q}}).
\end{equation}
When both pseudospin flavors have the same hopping integral, i.e.,
$t_{\uparrow }=t_{\downarrow }=t$, $H_t$ has the SU(2) symmetry,
leading to a vanishing $M_{{\bf kq}}$ and a free hole propagation with dispersion $%
-zt\gamma _{{\bf k}}$. Hence, the fact that $t_{\uparrow }\neq
t_{\downarrow }$ accounts for the formation of the
pseudospin-polaron even in the presence of a ferromagnetic
pseudospin background.

Using the self-consistent Born approximation (SCBA) in which the
spectral functions of one hole in a $t$-$J$-$like$ model can be
accurately calculated \cite
{HTC:SVR,yin:prl98,yin:prl01,CMR:bala}, we compute the
hole Green's function $G({\bf %
k},\omega )=[\omega -\varepsilon _{{\bf k}}^{}-\Sigma ({\bf
k},\omega )+i0^{+}]^{-1}$ self-consistently with the self-energy,
\begin{equation}
\Sigma ({\bf k},\omega )=\sum_{{\bf q}}M_{{\bf kq}}^2G({\bf k}-{\bf q}%
,\omega -\omega _{{\bf q}}). \label{se}
\end{equation}
Thus, the spectral functions are given by $A({\bf k},\omega
)=-\mathrm{Im} G({\bf k},\omega )/\pi $, and the quasiparticle
dispersion is $E_{{\bf k}}\equiv \varepsilon _{{\bf k}}+{\rm
Re}\Sigma ({\bf k},E_{{\bf k}})$.

{\em Antiferroelectric phase.} This phase
[Fig~\ref{fig:phases}(d)] is possible when $J_{\perp }>0$.
We find that linear spin wave theories of the antiferroelectric
and ferroelectric canting states are formally
connected with a displacement of momentum: ${\bf q}\rightarrow {\bf q}-%
{\bf Q}$ where ${\bf Q=}(\pi ,\ldots, \pi  )$. The effective
Hamiltonian for hole hopping is
\begin{equation}
H_t=\sum_{{\bf k}}\varepsilon _{{\bf k}}^{}h_{{\bf k}}^{\dagger }h_{{\bf k}%
}^{}+\sum_{{\bf k},{\bf q}}M_{{\bf kq}}h_{{\bf k}}^{\dagger }h_{{\bf k}-{\bf %
q}}^{}\alpha _{{\bf q-Q}}^{}+H.c.,
\end{equation}
where $\varepsilon _{{\bf k}}^{}=-z(t_{\downarrow }\cos ^2\frac \theta 2%
-t_{\uparrow }\sin ^2\frac \theta 2)\gamma _{{\bf k}}$ and
\begin{equation}
M_{{\bf kq}}=-\sin \theta \frac{t_{\uparrow }+t_{\downarrow }}2\frac z{\sqrt{%
N}}(\gamma _{{\bf k}-{\bf q}}u_{{\bf q-Q}}+\gamma _{{\bf
k}}v_{{\bf q-Q}}).
\end{equation}
Likewise, we employ (\ref{se}) to calculate the hole self energy.

{\em Staggered orbital ordering.} The pseudospin configuration is a $z$%
-directional N\'{e}el state [Fig 1(b)]. Considering the N\'{e}%
el state as the vacuum state, we define holon operators $%
h_{{\bf i}}$ and $g_{{\bf i}}$ so that $\,\,\widetilde{c}_{{\bf
i}\downarrow
}=h_{{\bf i}}^{\dagger },\,\,\,\widetilde{c}_{{\bf i}\uparrow }=h_{{\bf i}%
}^{\dagger }a_{{\bf i}}^{}$ on the $\downarrow $ sublattice and $\widetilde{%
\,c}_{{\bf j}\downarrow }=g_{{\bf j}}^{\dagger }b_{{\bf j}}^{},\,\,\,%
\widetilde{c}_{{\bf j}\uparrow }=g_{{\bf j}}^{\dagger }$ on the
$\uparrow $ sublattice. Here $a_{{\bf i}}^{}=\tau \,_{{\bf
i}}^{-}$ on the $\downarrow $ sublattice and $b_{{\bf j}}^{}=\tau
\,_{{\bf j}}^{+}$ on the $\uparrow $ sublattice are hard-core
boson operators. We arrive at an effective pseudospin-polaron
Hamiltonian
\begin{eqnarray}
H_t &=&\sum_{{\bf k},{\bf q}}{}^{\prime }M_{{\bf kq}}h_{{\bf
k}}^{\dagger }g_{{\bf k}-{\bf q}}^{}\alpha _{{\bf q}}+L_{{\bf
kq}}g_{{\bf k}}^{\dagger }h_{{\bf k}-{\bf q}}^{}\beta _{{\bf
q}}^{}+H.c.,
\end{eqnarray}
where $\alpha _{{\bf q}}$ and $\beta _{{\bf q}}$ are pseudospin
wave
operators, $a_{{\bf q}}=u_{{\bf q}}\alpha _{{\bf q}}+v_{{\bf q}}\beta _{-%
{\bf q}}^{\dagger }$ and$\;b_{-{\bf q}}=u_{{\bf q}}\beta _{-{\bf q}}+v_{{\bf %
q}}\alpha _{{\bf q}}^{\dagger }$, $\,$with dispersion $\omega _{{\bf q}%
}^{\pm }=zS(J_z^2-J_{\perp }^2\gamma _{{\bf q}}^2)^{1/2}\pm B_z$,
respectively, and $u_{{\bf q}%
}=\{[zSJ_z/(\omega _{{\bf q}}^{+}+\omega _{{\bf q}}^{-})+1]/2\}^{1/2}$, $v_{%
{\bf q}}=-{\rm sgn}(J_{\perp }\gamma _{{\bf q}})\{[zSJ_z/(\omega _{{\bf q}%
}^{+}+\omega _{{\bf q}}^{-})-1]/2\}^{1/2}$.
The hole-pseudospin-wave coupling is $M_{{\bf kq}}=-z%
\sqrt{2/N}(t_{\uparrow }\gamma _{{\bf k}-{\bf q}}u_{{\bf
q}}+t_{\downarrow }\gamma _{{\bf k}}v_{{\bf q}})$ and $L_{{\bf
kq}}=-z\sqrt{2/N}(t_{\downarrow
}\gamma _{{\bf k}-{\bf q}}u_{{\bf q}}+t_{\uparrow }\gamma _{{\bf k}}v_{{\bf q%
}})$. The summations over ${\bf k}$ and ${\bf q}$ are restricted
inside the reciprocal Brillouin zone of one sublattice. Within the
SCBA, we obtain the following self-consistent equations for the
two types of hole propagators: $G_j({\bf k},\omega )=[\omega
-\Sigma _j({\bf k},\omega )+i0^{+}]^{-1}$ where $j=g,h$ with
$\Sigma _h({\bf k},\omega )=\sum_{{\bf q}}{}^{\prime
}M_{{\bf kq}}^2G_g({\bf k}-{\bf q},\omega -\omega _{{\bf q}}^{+})$ and $%
\Sigma _g({\bf k},\omega )=\sum_{{\bf q}}{}^{\prime }L_{{\bf kq}}^2G_h({\bf k%
}-{\bf q},\omega -\omega _{{\bf q}}^{-})$.

To facilitate EFE, the system must be in a mix-valence regime and
the two bands involved should have different parity, thus
$t_{\uparrow }$ and $t_{\downarrow }$ have opposite signs, so do
$J_z$ and $J_{\perp }$; furthermore, it would be more probable if
both bands have similar bandwidths \cite{efe:batista}. Hence, we
adopt the following parameters \cite{note:parameter}:
$t_{\uparrow }=1$, $t_{\downarrow }=-0.8$, $U^{fd}=10$, hence $%
J_z=2(t_{\uparrow }^2+t_{\downarrow }^2)/U^{fd}=0.328$ and
$J_{\perp }=4t_{\uparrow }t_{\downarrow }/U^{fd}=-0.32$, with $%
B_z=\epsilon _d-\epsilon _f$ being a free parameter -- in a real
material, changing $B_z$ could be achieved by applying pressure or
alloying. Our numerical results are illustrated in the
two-dimensional case. In this case, the system transits from
staggered orbital ordering to the BEC of $d$-$f$ excitons at
$|B_z|=0.144$ and then to a non-mixed-valence regime at
$|B_z|=1.296$ as $B_z$ increases. All numerical calculations were
carried out on a $16\times 16$ square lattice.

\begin{figure}
\includegraphics*{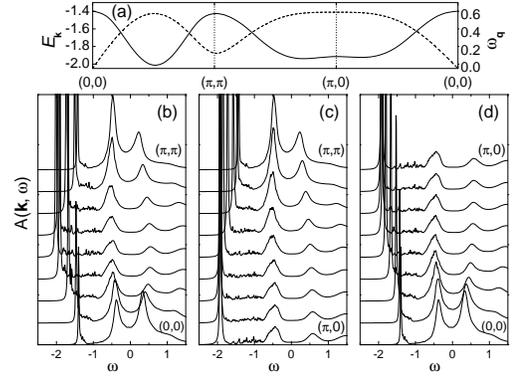}%
\caption{\label{fig:B02}%
(a) presents the quasiparticle dispersion $E_\mathbf{k}$ (solid
line, left scale), and pseudospin wave dispersion
$\omega_\mathbf{q}$ (dashed line, right scale) in the
ferroelectric phase with $B_z=0.2$. (b)-(d) shows the hole
spectral functions $A(\mathbf{k},\omega)$ along (b)
$(0,0)-(\pi,\pi)$, (c) $(\pi,\pi)-(\pi,0)$ and (d) $(\pi,0)-(0,0)$
directions.}
\end{figure}

\begin{figure}
\includegraphics*{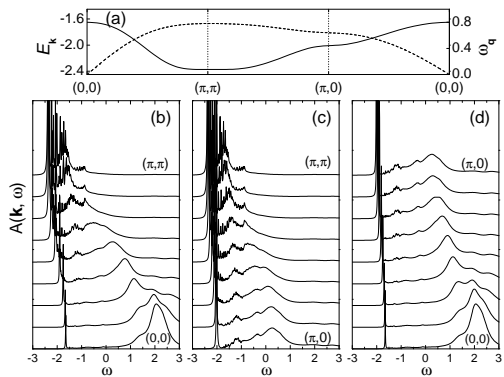}%
\caption{\label{fig:B08}%
The same quantities as in Fig.~\ref{fig:B02} but with $B_z=0.8$.}
\end{figure}

\begin{figure}
\includegraphics*{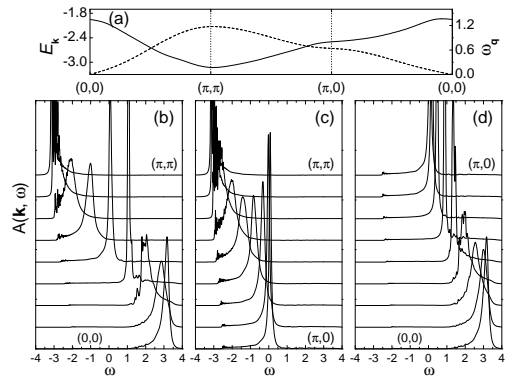}%
\caption{\label{fig:B12}%
The same quantities as in Fig.~\ref{fig:B02} but with $B_z=1.2$.}
\end{figure}

In Figs.~\ref{fig:B02}, \ref{fig:B08}, and \ref{fig:B12}, we show
the spectral functions and the quasiparticle dispersions in the
ferroelectric regime with $B_z=0.2$, $0.8$, and $1.2$,
respectively \cite{note:soo}. First of all, as presented in
Figs.~\ref{fig:B02}(a), \ref{fig:B08}(a), and \ref{fig:B12}(a),
the quasiparticle bandwidth $W$ and the pseudospin wave bandwidth
$W_\mathrm{spw}$ have similar values. Such band narrowing can be
understood in the following way: gapless pseudospin excitations
are easily stimulated by incoherent hole motion, leading to the
formation of the quasiparticle (QP), pseudospin polaron, which is
the propagating hole surrounded by a cloud of polarized pseudospin
waves. Therefore, $W$ does not scale with hopping integrals but
with $W_\mathrm{spw}$.

As displayed in Figs.~\ref{fig:B02}(b)-(d), for any $%
{\bf k}$, there is a well-defined quasiparticle pole (i.e., zero
pseudospin wave) at the low energy side which is well separated
from a broad, incoherent, multiple-pseudospin-wave background
extending to the full free-electron bandwidth.
Fig.~\ref{fig:B02}(a) shows that the bottom and the top of the QP
band locate at $(\pi/2,\pi/2)$ and $(0,0)$, respectively.
Furthermore, there is a flat region around $(\pi,0)$ at which the
QP energy is close to its minimum, leading to a strongly distorted
density of states with a massive peak near the bottom of the QP
band. These features obtained for single hole motion in the
ferromagnetic pseudospin background are quite similar to those in
the cuprate $t$-$J$ model where the hopping integrals for both
spin flavors are the same and the spin background is a Heisenberg
antiferromagnet \cite{yin:prl98}.
Note that the pseudospin wave spectrum does not vanish at
$(\pi,\pi)$ while it is gapless at $(0,0)$. Consequently, the QP
energy at $(\pi,\pi)$ is lower than that at $(0,0)$.

Further increasing $B_z$ will continue lowering the QP energy at
$(\pi,\pi)$. When $B_z=0.8$, $(\pi,\pi)$ becomes the QP band
bottom, as shown in Fig.~\ref{fig:B08}(a). While the flat region
around $(\pi,0)$ is shrinking, a new, surprising flat region
appears around $(\pi,\pi)$. Therefore, the feature of an extended
van Hove singularity near the bottom of the QP band survives up to
$B_z=0.8$. In Figs.~\ref{fig:B08}(b)-(d), the spectral functions
are still characterized by sharp peaks at the low energy side
which is well separated from a broad, incoherent background.
However, apart from the QP band bottom, considerable spectral
weights move to the incoherent part, indicating that the hole
motion is more severely damped by incoherent processes.

When $B_z$ is increased to $1.2$ [Figs.~\ref{fig:B12}(b)-(d)],
almost all spectral weights locate in the incoherent part except
near the QP band bottom, $(\pi,\pi)$. This means that
the hole spectral weights $Z({\bf k})=[1-\partial \Sigma (%
{\bf k},\omega )/\partial \omega ]_{\omega=E_\mathbf{k}}^{-1}$ are
strongly reduced by a cloud of polarized pseudospin waves. The
shape of the QP dispersion [Fig.~\ref{fig:B12}(a)] becomes similar
to that of free hole dispersion $\propto \gamma_\mathbf{k}$, yet
with bandwidth $W \simeq W_\mathrm{spw}$. As $B_z$ further
increases, $Z(\mathbf{k})$ at wave vectors not very close to
$(\pi,\pi)$ goes to zero quickly until the pseudospin polaron
picture becomes invalid at $B_z=1.296$. Then, the system evolves
into full $f$-band with free hole motion, implying that
$Z(\mathbf{k})\equiv 1$.

To summarize, we present a systematic study on the evolution of
the spectral functions of a single hole in the FKM as a function
of $B_z$, the $d$- and $f$-level energy difference. We find that
strong scattering processes between the hole and $d$-$f$ excitons
exist in the mix-valence regimes, thus lead to the formation of
pseudospin polarons and interesting electronic structures with
bandwidth scaling with that of the pseudospin excitation, even for
single hole propagation in a ferromagnetic pseudospin background
(i.e., the ferroelectric phase) due to the different hopping
integrals of the $d$- and $f$-electrons. Furthermore, flat regions
near the bottom of the quasiparticle band are found in a wide
range of $B_z$ in the two-dimensional case. As we know, similar
anomalous, yet crucial hole spectral features were found in high
temperature superconductors \cite{HTC:dagotto2,yin:prl98}, thus we
speculate that besides EFE, the extended Falicov-Kimball model
could have many novel properties, especially upon hole doping.


\begin{acknowledgments}
This work is supported by the Nebraska Research Initiative, the
Nebraska EPSCOR-NSF Grant EPS-9720643, and Department of the Army
Grants DAAG 55-98-1-0273 and DAAG 55-99-1-0106. W. N. M. is
grateful for the support from the Office of Naval Research.
\end{acknowledgments}



\begin{thebibliography}{17}
\expandafter\ifx\csname
natexlab\endcsname\relax\def\natexlab#1{#1}\fi
\expandafter\ifx\csname bibnamefont\endcsname\relax
  \def\bibnamefont#1{#1}\fi
\expandafter\ifx\csname bibfnamefont\endcsname\relax
  \def\bibfnamefont#1{#1}\fi
\expandafter\ifx\csname citenamefont\endcsname\relax
  \def\citenamefont#1{#1}\fi
\expandafter\ifx\csname url\endcsname\relax
  \def\url#1{\texttt{#1}}\fi
\expandafter\ifx\csname
urlprefix\endcsname\relax\def\urlprefix{URL }\fi
\providecommand{\bibinfo}[2]{#2}
\providecommand{\eprint}[2][]{\url{#2}}

\bibitem[{\citenamefont{Kittel}(1996)}]{efe:kittel}
\bibinfo{author}{\bibfnamefont{C.}~\bibnamefont{Kittel}},
  \emph{\bibinfo{title}{Introduction to Solid State Physics}}
  (\bibinfo{publisher}{John Wiely and Sons, Inc.}, \bibinfo{address}{New York},
  \bibinfo{year}{1996}).

\bibitem[{\citenamefont{Portengen
  et~al.}(1996{\natexlab{a}})\citenamefont{Portengen, \"{O}streich, and
  Sham}}]{efe:portengen}
\bibinfo{author}{\bibfnamefont{T.}~\bibnamefont{Portengen}},
  \bibinfo{author}{\bibfnamefont{T.}~\bibnamefont{\"{O}streich}},
  \bibnamefont{and} \bibinfo{author}{\bibfnamefont{L.~J.} \bibnamefont{Sham}},
  \bibinfo{journal}{Phys.\ Rev.\ Lett.} \textbf{\bibinfo{volume}{76}},
  \bibinfo{pages}{3384} (\bibinfo{year}{1996}{\natexlab{a}}).

\bibitem[{\citenamefont{Portengen
  et~al.}(1996{\natexlab{b}})\citenamefont{Portengen, \"{O}streich, and
  Sham}}]{efe:portengen-prb}
\bibinfo{author}{\bibfnamefont{T.}~\bibnamefont{Portengen}},
  \bibinfo{author}{\bibfnamefont{T.}~\bibnamefont{\"{O}streich}},
  \bibnamefont{and} \bibinfo{author}{\bibfnamefont{L.~J.} \bibnamefont{Sham}},
  \bibinfo{journal}{Phys.\ Rev.\ B} \textbf{\bibinfo{volume}{54}},
  \bibinfo{pages}{17452} (\bibinfo{year}{1996}{\natexlab{b}}).

\bibitem[{\citenamefont{Falicov and Kimball}(1969)}]{efe:FK}
\bibinfo{author}{\bibfnamefont{L.~M.} \bibnamefont{Falicov}} \bibnamefont{and}
  \bibinfo{author}{\bibfnamefont{J.~C.} \bibnamefont{Kimball}},
  \bibinfo{journal}{Phys.\ Rev.\ Lett.} \textbf{\bibinfo{volume}{22}},
  \bibinfo{pages}{997} (\bibinfo{year}{1969}).

\bibitem[{\citenamefont{Czycholl}(1999)}]{efe:czycholl}
\bibinfo{author}{\bibfnamefont{G.}~\bibnamefont{Czycholl}},
  \bibinfo{journal}{Phys.\ Rev.\ B} \textbf{\bibinfo{volume}{59}},
  \bibinfo{pages}{2642} (\bibinfo{year}{1999}).

\bibitem[{\citenamefont{Farka\v{s}ovsk\'{y}}(2002)}]{efe:farkasovsky-02}
\bibinfo{author}{\bibfnamefont{P.}~\bibnamefont{Farka\v{s}ovsk\'{y}}},
  \bibinfo{journal}{Phys.\ Rev.\ B} \textbf{\bibinfo{volume}{65}},
  \bibinfo{pages}{081102(R)} (\bibinfo{year}{2002}).

\bibitem[{\citenamefont{Zlati\'{c}}(2001)}]{efe:zlatic}
\bibinfo{author}{\bibfnamefont{V.}~\bibnamefont{Zlati\'{c}}},
  \bibinfo{journal}{Philos. Mag. B} \textbf{\bibinfo{volume}{81}},
  \bibinfo{pages}{1443} (\bibinfo{year}{2001}).

\bibitem[{\citenamefont{Batista}(2002)}]{efe:batista}
\bibinfo{author}{\bibfnamefont{C.~D.} \bibnamefont{Batista}},
  \bibinfo{journal}{Phys.\ Rev.\ Lett.} \textbf{\bibinfo{volume}{89}},
  \bibinfo{pages}{166403} (\bibinfo{year}{2002}).

\bibitem[{\citenamefont{Schmid et~al.}(2002)\citenamefont{Schmid, Todo, Troyer,
  and Dorneich}}]{efe:schmid}
\bibinfo{author}{\bibfnamefont{G.}~\bibnamefont{Schmid}},
  \bibinfo{author}{\bibfnamefont{S.}~\bibnamefont{Todo}},
  \bibinfo{author}{\bibfnamefont{M.}~\bibnamefont{Troyer}}, \bibnamefont{and}
  \bibinfo{author}{\bibfnamefont{A.}~\bibnamefont{Dorneich}},
  \bibinfo{journal}{Phys.\ Rev.\ Lett.} \textbf{\bibinfo{volume}{88}},
  \bibinfo{pages}{167208} (\bibinfo{year}{2002}).

\bibitem[{\citenamefont{Dagotto et~al.}(1995)\citenamefont{Dagotto, Nazarenko,
  and Moreo}}]{HTC:dagotto2}
\bibinfo{author}{\bibfnamefont{E.}~\bibnamefont{Dagotto}},
  \bibinfo{author}{\bibfnamefont{A.}~\bibnamefont{Nazarenko}},
  \bibnamefont{and} \bibinfo{author}{\bibfnamefont{A.}~\bibnamefont{Moreo}},
  \bibinfo{journal}{Phys.\ Rev.\ Lett.} \textbf{\bibinfo{volume}{74}},
  \bibinfo{pages}{310} (\bibinfo{year}{1995}).

\bibitem[{\citenamefont{Schmitt-Rink et~al.}(1988)\citenamefont{Schmitt-Rink,
  Varma, and Ruckenstein}}]{HTC:SVR}
\bibinfo{author}{\bibfnamefont{S.}~\bibnamefont{Schmitt-Rink}},
  \bibinfo{author}{\bibfnamefont{C.~M.} \bibnamefont{Varma}}, \bibnamefont{and}
  \bibinfo{author}{\bibfnamefont{A.~E.} \bibnamefont{Ruckenstein}},
  \bibinfo{journal}{Phys.\ Rev.\ Lett.} \textbf{\bibinfo{volume}{60}},
  \bibinfo{pages}{2793} (\bibinfo{year}{1988}).

\bibitem[{\citenamefont{Yin et~al.}(1998)\citenamefont{Yin, Gong, and
  Leung}}]{yin:prl98}
\bibinfo{author}{\bibfnamefont{W.-G.} \bibnamefont{Yin}},
  \bibinfo{author}{\bibfnamefont{C.~D.} \bibnamefont{Gong}}, \bibnamefont{and}
  \bibinfo{author}{\bibfnamefont{P.~W.} \bibnamefont{Leung}},
  \bibinfo{journal}{Phys.\ Rev.\ Lett.} \textbf{\bibinfo{volume}{81}},
  \bibinfo{pages}{2534} (\bibinfo{year}{1998}).

\bibitem[{\citenamefont{Yin et~al.}(2001)\citenamefont{Yin, Lin, and
  Gong}}]{yin:prl01}
\bibinfo{author}{\bibfnamefont{W.-G.} \bibnamefont{Yin}},
  \bibinfo{author}{\bibfnamefont{H.~Q.} \bibnamefont{Lin}}, \bibnamefont{and}
  \bibinfo{author}{\bibfnamefont{C.~D.} \bibnamefont{Gong}},
  \bibinfo{journal}{Phys.\ Rev.\ Lett.} \textbf{\bibinfo{volume}{87}},
  \bibinfo{pages}{047204} (\bibinfo{year}{2001}).

\bibitem[{\citenamefont{Bala et~al.}(2001)\citenamefont{Bala, Sawatzky,
  Ole\'{s}, and Macridin}}]{CMR:bala}
\bibinfo{author}{\bibfnamefont{J.}~\bibnamefont{Bala}},
  \bibinfo{author}{\bibfnamefont{G.~A.} \bibnamefont{Sawatzky}},
  \bibinfo{author}{\bibfnamefont{A.~M.} \bibnamefont{Ole\'{s}}},
  \bibnamefont{and} \bibinfo{author}{\bibfnamefont{A.}~\bibnamefont{Macridin}},
  \bibinfo{journal}{Phys.\ Rev.\ Lett.} \textbf{\bibinfo{volume}{87}},
  \bibinfo{pages}{067204} (\bibinfo{year}{2001}).

\bibitem[{not({\natexlab{a}})}]{note:phases}
\bibinfo{note}{The phase diagram of $H_J$ is symmetric under a change of sign
  of $B_z$ because $H_J$ is symmetric under a reflection in the $xy$ plane}.

\bibitem[{not({\natexlab{b}})}]{note:parameter}
\bibinfo{note}{These parameters are close to those for cuprates and manganites,
  see Ref. \cite{yin:prl98,yin:prl01}}.

\bibitem[{not({\natexlab{c}})}]{note:soo}
\bibinfo{note}{The results in staggered orbital ordering are similar to those
  in the ferroelectric phase with $B_z=0.2$}.

\end{thebibliography}

\end{document}